\documentclass[a4paper,12pt]{article}

\begin{document}
\makeatletter
\renewcommand{\theequation}{\thesection.\arabic{equation}}
\@addtoreset{equation}{section}
\makeatother

\title{\Large Contracted Representation of Yang's Space-Time Algebra and Buniy-Hsu-Zee's Discrete 
Space-Time    }
\author{\large Sho Tanaka\footnote{Em.~Professor of Kyoto University, E-mail: stanaka@yukawa.kyoto-u.ac.jp, 
 Kurodani 33-4, Sakyo-ku, Kyoto 606-8331, Japan.}}
\date{}
\maketitle

\vspace{-10cm}
\rightline{}
\vspace{12cm}
\thispagestyle{empty}

Motivated by the recent proposition by Buniy, Hsu and Zee with respect to discrete space-time 
and finite spatial degrees of freedom of our physical world with a short- and a long-distance scales, 
$l_P$ and $L,$ we reconsider the Lorentz-covariant Yang's quantized space-time algebra (YSTA), which 
is intrinsically equipped with such two kinds of scale parameters, $\lambda$ and $R$. In accordance with 
their proposition, we find the so-called contracted representation of YSTA with finite spatial degrees of 
freedom associated with the ratio $R/\lambda$, which gives a possibility of the divergence-free 
noncommutative field theory on YSTA.  
The canonical commutation relations familiar in the ordinary quantum 
mechanics appear as the cooperative Inonu-Wigner's contraction limit of YSTA, $\lambda \rightarrow 0$ 
and $R \rightarrow \infty.$

\vspace{2\baselineskip}

\newpage

\section{\normalsize Introduction}
As was pointed out recently by Jackiw$^{[1]}$, the idea of current noncommutative quantized 
space-time was first suggested by Heisenberg in the late 1930's so as to regulate  the
short-distance singularities in local quantum-field theories by virtue of the noncommutative-coordinate 
uncertainty. In 1947, Snyder$^{[2]}$ worked out it successfully over a lot of challenges by prominent 
theoretical physicists in those days. Immediately after the Snyder's pioneering work,
Yang$^{[3]}$ proposed the so-called Yang's space-time algebra by introducing quantized space-time and 
momentum in parallel, in order to improve the Snyder's theory so as to satisfy the translation invariance in addition 
to the Lorentz-invariance, which nicely holds in both theories. Unfortunately, however, in those 
approaches$^{[4]}$
we have never succeeded so far to find a clear-cut conclusion on the original Heisenberg's intention to 
get rid of short-distance singularities, in spite of the fact that especially in the Yang's theory the 
so-called contraction parameters, $\lambda$ and $R$ are explicitly introduced which might be related to 
short-distance (ultraviolet) and long-distance (infrared) regularization, respectively, as will be 
shown later in the present paper.
    
    In this connection, the recent idea proposed by Buniy, Hsu and Zee$^{[5]}$, let us refer it hereafter 
by the B-H-Z's 
proposition, is noticeable. Let us summarize its essence according to our present viewpoint: (1) There are 
many indications that space-time may be discrete rather than continuous. A consequence of spatial 
discreteness with spacing $l_P$ is that in any finite region of size $L$ there are only a finite number of 
degrees of freedom $N \sim (L/l_P)^3.$ (2) Although our universe might be infinite in extent, any experiment 
performed by scientists must take place over a finite period of time. By causality, this implies that 
the experiment takes place in a region of finite size, which we take to be $L$. We therefore assume the 
existence of a long-distance (infrared) regulator $L$ in addition to a short-distance (ultraviolet) 
regulator $l_P$.

 Indeed, it is well known that a reasonable 
derivation of a finite number of spatial degrees of freedom is crucial for arriving at divergence-free 
quantum field theory within an ordinary Hilbert space of countable degrees of freedom as in quantum 
mechanics. The problem will be our main subject in this paper.

The present author has long studied the Snyder-Yang's quantized space-time, 
through a series of the preliminary studies$^{[6,7]}$, since the first proposal$^{[8]}$
of Snyder-Yang's quantized space-time applied to the matrix model$^{[9]}$. Especially, 
the so-called Yang's quantized space-time algebra has been there noticed, 
because of the fact that it is intrinsically equipped with two scale parameters, $\lambda$ and $R$, deeply 
related to $l_P$ and $L$ in the above B-H-Z's proposition and has the background symmetry equivalent to 
a conformal symmetry$^{[6]}$. As will be shown in the present paper, it has 
a strong possibility of leading us to a noncommutative field theory free from the ultraviolet divergences  
with aid of the B-H-Z's proposition. 

The present paper is organized as follows. In Sec. 2, we shortly recapitulate Yang's space-time algebra 
(YSTA) and its spatial discrete structure in connection with the short-distance parameter in the 
B-H-Z's proposition. In Sec. 3, based on the consideration of the quantum-mechanical limit of YSTA, we 
present a 
postulate of a certain contraction of representation space 
of YSTA in connection with the long-distance scale in the B-H-Z's 
proposition. Sec. 4 is devoted to the group-theoretical consideration of our postulate of contracted 
representation of YSTA with finite spatial degrees of freedom. In Sec. 5, we show a possibility of divergence-free noncommutative field theory 
on YSTA. In the final section, Sec. 6, the physical and theoretical 
implication of our postulate and the B-H-Z's proposition is conclusively discussed.
\vspace{1\baselineskip}
\section{\normalsize Yang's Space-Time Algebra (YSTA) and B-H-Z's Proposition}
    Let us first review the essence of Yang's space-time algebra, focusing our attention on the B-H-Z's 
proposition$^{[5]}$.

    As was explicitly shown in Ref. [7],  $D$-dimensional Yang's quantized space-time algebra (YSTA) is derived
as the result of the so-called Inonu-Wigner$^{[10]}$ contraction procedure with two contraction parameters, 
$R$ and $\lambda$, from $SO(D+1,1)$ algebra with 
generators 
$\hat{\Sigma}_{MN}$; 
\begin{eqnarray}
 \hat{\Sigma}_{MN}  \equiv i (q_M \partial /{\partial{q_N}}-q_N\partial/{\partial{q_M}}),
\end{eqnarray}
which work on $(D+2)$-dimensional parameter space  $q_M$ ($M= \mu,a,b)$ satisfying  
\begin{eqnarray}
             - q_0^2 + q_1^2 + ... + q_{D-1}^2 + q_a^2 + q_b^2 = R^2.
\end{eqnarray}
 
Here, $q_0 =-i q_D$ and $M = a, b$ denote two extra dimensions with space-like metric signature.

$D$-dimensional space-time and momentum operators, $\hat{X}_\mu$ and $\hat{P}_\mu$, 
with $\mu =1,2,...,D,$ are defined in parallel by
\begin{eqnarray}
&&\hat{X}_\mu \equiv \lambda\ \hat{\Sigma}_{\mu a}
\\
&&\hat{P}_\mu \equiv \hbar /R \ \hat{\Sigma}_{\mu b},   
\end{eqnarray}
together with $D$-dimensional angular momentum operator $\hat{M}_{\mu \nu}$
\begin{eqnarray}
   \hat{M}_{\mu \nu} \equiv \hbar \hat{\Sigma}_{\mu \nu}
\end{eqnarray} 
and the so-called reciprocity operator
\begin{eqnarray}
    \hat{N}\equiv \lambda /R\ \hat{\Sigma}_{ab}.
\end{eqnarray}

Operators  $( \hat{X}_\mu, \hat{P}_\mu, \hat{M}_{\mu \nu}, \hat{N} )$ defined above 
satisfy the so-called contracted algebra of the original $SO(D+1,1)$, or Yang's 
space-time algebra (YSTA):
\begin{eqnarray}
&&[ \hat{X}_\mu, \hat{X}_\nu ] = - i \lambda^2/\hbar \hat{M}_{\mu \nu}
\\
&&[\hat{P}_\mu,\hat{P}_\nu ] = - i\hbar / R^2\ \hat{M}_{\mu \nu}
\\
&&[\hat{X}_\mu, \hat{P}_\nu ] = - i \hbar \hat{N} \delta_{\mu \nu}
\\
&&[ \hat{N}, \hat{X}_\mu ] = - i \lambda^2 /\hbar  \hat{P}_\mu
\\
&&[ \hat{N}, \hat{P}_\mu ] =  i \hbar/ R^2\ \hat{X}_\mu,
\end{eqnarray}
with familiar relations among ${\hat{M}_{\mu \nu}}\ 's$ omitted.

Here, one finds that the so-called contraction-parameters $\lambda$ and $R$ with dimension of length 
introduced in the above expression are fundamental constants of YSTA, which are to be compared with 
two fundamental scales of short-distance $l_P$ and long-distance $L$, respectively, in the B-H-Z's 
proposition mentioned in the preceding section.  

Before entering into the central problem in the present paper, it is important to notice 
the following basic fact that ${\hat\Sigma}_{MN}$ defined in Eq. (2.1) with $M, N$ being the same metric 
signature have discrete eigenvalues, i.e., $0,\pm 1 ,\pm 2,\dots$, and those with $M, N$ being opposite 
metric signature have continuous eigenvalues, as was shown in Ref. [7] explicitly. This fact leads  
to the remarkable result of YSTA: The spatial components of position and momentum defined in Eqs. (2.3)
and (2.4), respectively, have discrete eigenvalues in units of $\lambda$ and $\hbar/R$, and temporal components of them 
continuous eigenvalues, consistently with Lorentz invariance. As was emphasized by 
Yang and explicitly shown in Ref. [7], this conspicuous fact is entirely due to noncommutativity 
between individual components of space-time and momentum operators. This aspect of YSTA is 
well understood by means of the familiar example of the three-dimensional angular momentum in 
quantum mechanics, where individual components, which are noncommutative among themselves, are able 
to have discrete eigenvalues, consistently with the three-dimensional rotation-invariance.

It should be noted that in YSTA the requirement of spatial discreteness in the B-H-Z's proposition 
is well satisfied in a Lorentz-covariant way and the fundamental scale parameter $\lambda$ actually 
has the physical meaning as the minimal length of spatial distance in Eq. (2.3). With respect to 
another fundamental scale parameter $R$, however, the preceding argument indicates only the fact that it 
gives the minimal scale of momenta, $\hbar/R$. As a matter of fact, Eq. (2.3), as it stands, tells us that the 
maximal eigenvalue of spatial 
components of position operators $\hat X_i$ becomes infinite corresponding to the large limit of integer 
eigenvalue of $\hat \Sigma_{ia}$. In consequence, one finds that this is the central problem of 
YSTA in arriving at the B-H-Z's proposition.
\vspace{1\baselineskip}
\section{\normalsize Quantum-Mechanical Limit of YSTA and Contraction of Representation Space}

It is important here to notice that a clue to this problem is found in our previous 
consideration$^{[7]}$ on the quantum-mechanical limit of YSTA, in which the canonical commutation 
relations familiar in the ordinary quantum mechanics appear. This consideration was given in 
the course of investigating the translation 
operation, which was introduced by Yang beyond the original Snyder's space-time algebra, as mentioned above. 
Let us here review it shortly.

In fact, D-dimensional translation operator ${\hat T}$ with 
infinitesimal parameters $\alpha_\mu$ is defined by
\begin{eqnarray}
          {\hat T}(\alpha_\mu) = \exp\ i\ (\alpha_\mu\ {\hat P}_\mu)\
( =  \exp\ ({i\hbar {\alpha}_\mu \over R}{\hat \Sigma}_{b\mu} )).
\end{eqnarray}
One finds that this operator induces infinitesimal transformation on $\hat{X}_\mu$ 
\begin{eqnarray}
             \hat{X}_\mu \rightarrow \hat{X}_\mu + \alpha_\mu\ \hat{N}
\end{eqnarray}
together with
\begin{eqnarray}
               \hat{N} \rightarrow \hat{N} -  \alpha_\mu\ \hat{X}_\mu / R^2.
\end{eqnarray}
This result is well understood, if one notices that the momentum operator ${\hat P}_\mu$ in YSTA is 
nothing but generator of infinitesimal rotation on $b-\mu$ plane, ${\hat \Sigma}_{\mu b}$ given 
in Eq. (2.4).

However, let us here notice that the reciprocity operator $\hat{N} (= \lambda R^{-1} 
\hat{\Sigma}_{ab})$ defined in Eq. (2.6) is an operator with discrete eigenvalues $ n\ 
(\lambda R^{-1})$, $n$ being $\pm$ integer and the displacement $ \alpha_\mu \hat{N}$ in Eq. (3.2) is 
noncommutative with ${\hat X}_\mu$.  Therefore, it is important to see in what limit the ordinary 
translations familiar in quantum mechanics may appear. Indeed, one finds them in a following 
cooperative limit of contraction-parameters, $\lambda$ and $R$
\begin{eqnarray}
\lambda &\rightarrow& 0 
\nonumber\\
R &\rightarrow& \infty,
\end{eqnarray} 
in conformity with a condition
\begin{eqnarray}
     \hat N (= \lambda R^{-1} {\hat \Sigma}_{ab}) \rightarrow 1.
\end{eqnarray}

In fact, the above condition (3.5) in the limit (3.4) can never be a simple operator relation of 
YSTA and turns out below to be of the so-called Inonu-Wigner 
``contraction of groups and their representations"$^{[10]}$. One finds immediately that it necessitates a large 
limit of 
discrete eigenvalues of ${\hat\Sigma}_{ab}$\ in order for $\hat N$\ to survive with nonvanishing value $1$. 

Furthermore, it should be noted here that the canonical commutation relations in the ordinary quantum 
mechanics just appear as the above cooperative 
Inonu-Wigner contraction limit of YSTA with respect to $\lambda$ and $R$, as seen from (2.7) to (2.11). 
This fact reminds us the Bohr's correspondence principle at the birth of quantum mechanics, that is, 
quantum mechanics tends to classical mechanics in a large limit of quantum numbers. Hereafter, let us 
call the limit of contraction parameters (3.4) under the constraint (3.5), the quantum-mechanical limit 
of YSTA.

At this point, let us rewrite the last Eq. (3.5) in the following form
\begin{eqnarray} 
    {\hat \Sigma}_{ab} \rightarrow R/\lambda\ (infinite).
\end{eqnarray}
As was remarked after Eq. (3.5), this limiting relation as well as (3.5) does not imply a simple operator 
relation of ${\hat \Sigma}_{ab}$ in YSTA, but seems strongly to suggest that in YSTA equipped with 
contraction parameters $R$ and $\lambda$ in accordance with the B-H-Z's proposition, eigenvalues of 
${\hat \Sigma}_{ab}$ or more covariantly eigenvalues $n_{MN}$ of generators 
${\hat \Sigma}_{MN}$ of SO(D+1), 
with $M, N =  a, b, 1, 2, \dots, D-1$, are to be limited 
to integers between  $+R/\lambda$ and $-R/\lambda$ with the largest integer $ [R/\lambda ]$, so that 
Eq. (3.6) is replaced by the more comprehensive expression
\begin{eqnarray}
        \vert n_{MN} \vert \leq  {\cal N}\ (\equiv [R/\lambda] ) \rightarrow infinite, \quad (M,N = a, b, 1, 2, \dots, 
D-1)
\end{eqnarray} 
with 
\begin{eqnarray}
      {\cal N} \equiv [R/\lambda].
\end{eqnarray}
 
In fact, this postulate applied to Eqs. (2.3) and (2.4), respectively, leads to the result that the contraction 
parameter $R$ indicates the maximal spatial scale in YSTA, as was expected in 
the B-H-Z's proposition, and the maximal eigenvalue of spatial components of momenta becomes 
$\hbar /\lambda$. 

The mathematical and physical implication of the postulate (3.7) will be clarified 
in the next Sec. 4. \footnote{Our postulate stated above suggests a certain 
possibility of a theoretical refinement of 
the Inonu-Wigner contraction scheme$^{[10]}$ applied in physics, in which the representation space of 
the contracted algebra with finite 
values of contraction parameters, might be so constrained that 
contraction parameters, not merely as mathematical ones, but acquire their own physical meanings, as was 
shown above in the case of YSTA.} 
\vspace{1\baselineskip}
\section{\normalsize Irreducible Decomposition of $SO(D+1)$-
Representation in Quasi-Regular Representation of YSTA 
$(SO(D+1,1))$ }

In this section, let us first give a mathematical consideration of our postulate given by Eq. (3.7) in the 
preceding section. In Ref. [7], we studied the unitary infinite-dimensional representation of YSTA, let us call it 
the quasi-regular representation of $SO(D+1,1)$$^{[11]}$, in which the (noncompact) time operator 
${\hat X}_{0} \sim {\hat \Sigma}_{0a} $ is diagonal with continuous eigenvalue $t$ 
and the representation bases at a fixed time $t$ are given by simultaneous eigenstates of operators, 
which are commutative with time operator ${\hat X}_0 (=\lambda {\hat \Sigma}_{0a})$ and commutative 
among themselves, that is, the maximal commuting set of $SO(D)$, for instance, in $D(=11)$-dimensional 
YSTA, ${\hat{\Sigma}}_{12}$, ${\hat{\Sigma}}_{34},$..., ${\hat{\Sigma}}_{9 10}$. The whole of the 
representation bases are then given by the following infinite set of eigenstates, neglecting their 
multiplicity:
\begin{eqnarray}
|\ t\ ; n_{12}, \dots, n_{9 10}> \equiv|\ {\hat{\Sigma}}_{0a}={t/\lambda}; 
 {\hat{\Sigma}}_{12} = n_{12}, \dots, {\hat{\Sigma}}_{9 10} = n_{9 10}>.
\end{eqnarray}

Our postulate stated in the preceding section, implies that the absolute values of eigenvalues $ n_{i j} $ in the above expression 
do not exceed ${\cal N} ( = [R/\lambda])$. In consequence, the spatial degrees of freedom of the 
representation space of YSTA, we called it Hilbert Space I in Refs. [6,7], at a fixed time 
turn out to be finite and of the order of
\begin{eqnarray}
                        {\cal D} \equiv {\cal N}^{[D/2]} (= [R/\lambda]^{[D/2]})
\end{eqnarray}
with $[D/2]$ denoting the integer either of $D/2$ and $(D-1)/2$.

Our preceding postulate, i.e., the idea of setting a certain maximum ${\cal N}( =[R/\lambda] )$ to 
the eigenvalues of ${\hat \Sigma}_{ab}$, more covariantly those of generators of $SO(D+1)$, ${\hat \Sigma}_{MN}$ 
with $M, N = a, b, 1, 2, \dots, D-1$ 
may be defined in a more refined mathematical expression. Indeed, the quasi-regular representation of 
$SO(D+1,1)$ mentioned above, whose basis vectors are given by Eq. (4.1), must provide a certain unitary 
infinite-dimensional representation of its maximal compact subgroup $SO(D+1)$ under consideration, which can be 
completely reducible 
into an infinite series of unboundedly increasing-dimensional unitary irreducible representation 
of $SO(D+1)$. Therefore, our idea of setting a certain upper limit to eigenvalues of generators of $SO(D+1)$ 
means now to contract the above infinite series of irreducible decomposition into the finite series.

 Let us explain the situation more in detail$^{[12]}$. The above unitary representation of $S(D+1)$ 
appearing in 
the quasi-regular representation of $SO(D+1,1)$ is now explicitly decomposed into  a series of the 
irreducible representations expressed in terms of the so-called spherical functions defined on $S^D = 
SO(D+1)/SO(D)$, by taking $SO(D)$ with generators ${\hat \Sigma}_{MN}$ with $M,N = b, 1, 2, \dots, D-1.$ 

It is well-known that the irreducible representation $\rho_l$ of 
$SO(D+1)$ on the representation space $S^D$, is uniquely designated by the maximal integer $l$ of 
eigenvalues $n_{ab}$ of ${\hat \Sigma}_{ab}$ in the representation, which is known to be a possible Cartan 
subalgebra of the so-called compact symmetric pair $(SO(D+1),SO(D))$ of rank $1$$^{[12]}$. The whole set of basis 
spherical 
functions belonging 
to the irreducible representation $\rho_l$ 
is given by
\begin{eqnarray}
 \Phi_{l n_{12} n_{34} n_{56}\dots }( \theta_1, \theta_2, \dots,
\theta_D) \quad( l \ge n_{12}\ge n_{34}\ge n_{56}\ge \dots).
\end{eqnarray}
They are labeled by $n_{12}, n_{34}, n_{56},\dots$, i.e., 
eigenvalues of generators, ${\hat{\Sigma}}_{12}, {\hat{\Sigma}}_{34}, {\hat{\Sigma}}_{56},$\break
..., respectively, which constitute 
together with ${\hat \Sigma}_{ab}$  a Cartan subalgebra of $SO(D+1)$, and expressed in terms of 
the products of associated Gegenbauer functions of polar coordinates of $S^D,\ 
\theta_1, \theta_2,\dots, \theta_D$. 

Eigenvalues of Casimir operator of $SO(D+1)$ and of Laplace-Beltrami operator ${\it \Delta}$ 
operating on the spherical functions (4.3), are equally given by
\begin{eqnarray}
              -\ l(l+D-1).
\end{eqnarray}    

 The dimension of the representation $\rho_l$ is given by

\begin{eqnarray}
         dim (\rho_l)= {(l+\nu) \over \nu} {(l+2\nu-1)! \over {l!(2\nu -1)!}},
\end{eqnarray}
where $ \nu \equiv (D-1)/2$ and $D \geq 2.$

From the above argument, one finds out that our postulate to impose a certain maximum 
${\cal N}( =[R/\lambda] )$ upon the eigenvalues of generators of $S(D+1)$, that is, ${\hat \Sigma}_{MN}$ with 
$M, N= a, b, 1,2,\dots, D-1,$ is nothing but to impose the upper limit upon the series of 
irreducible 
representations $\rho_l $'s of $SO(D+1)$ to be $ l \le {\cal N} ( =[R/\lambda])$.\footnote{It is 
interesting to find a certain similar procedure in Ref. [10], 
where a contraction limit of $SO(3)$ into the inhomogeneous two-dimensional rotation group is shown through 
the infinite sequence of irreducible representations of $SO(3)$.}  

At this point, it is important to reconsider the physical implication of our postulate to limit or contract 
the representation space of YSTA in accordance with the B-H-Z's proposition. In fact, it 
is clear that the 
Lorentz-covariant quasi-regular representation of YSTA or $SO(D+1,1)$, whose representation bases are given 
in Eq. (4.1), formally gives rise to the {\it infinite} series of unitary irreducible representations 
$\rho_l$ 's of $SO(D+1)$ which clearly involves $l$'s beyond ${\cal N} ( =[R/\lambda])$. However, according to 
the B-H-Z's proposition, as stated in Introduction, our physical experiments performed by scientists take 
place over a finite period of time, at most, 
of the order of $R/c$, $R$ being of a cosmological scale, so the experiments must take place, by causality, 
in a region of finite spatial size, at most, of the order of $R$. In other words, any processes, which 
might evolve beyond the contracted representation space of YSTA stated above, are to be regarded as 
unphysical and excluded in our present scheme tightly based on the B-H-Z's proposition, even though they are 
conceivable metaphysically. 

This means that although, as was mentioned above, the quasi-regular 
representation of $SO(D+1,1)$ formally provides the infinite series of irreducible decomposition of the 
unitary representation of its subgroup $SO(D+1)$, physical processes in our physical world necessitate and 
actually 
utilize for their own descriptions, the contracted part of representation space, which corresponds 
to the following order of {\it finite} spatial degrees of freedom
\begin{eqnarray}
  {\cal D}({\cal N}) = \sum_{l \le {\cal N}}\ dim \ (\rho_ l ),  
\end{eqnarray}
with ${\cal N}$ given in Eq. (3.8).

As seen in the next section, this consideration will play an important role in the arguments of ultraviolet 
divergence problem in the quantum field theory developed on YSTA.
\vspace{1\baselineskip}
\section{\normalsize Quantum Field Theory on YSTA with Finite Spatial Degrees of Freedom and B-H-Z's Proposition}

In a series of our studies of YSTA referred in the preceding argument, we have constantly tried to 
reformulate the so-called matrix model$^{[9]}$ as quantum mechanics of many-$D_0$ branes in terms of 
noncommutative quantum field theory on Yang's quantized space-time. One encounters there the concept 
of creation-annihilation operators of $D_0$ brane, which construct the Fock space of $D_0 $ branes, which 
we call Hilbert space II, distinguished from Hilbert space I so far discussed. 

At this point, it is very important to note that the number of different
creation-annihilation operators at fixed 
time $t$ is now not only countable but also {\it finite}, being of the order of ${\cal D({\cal N})}^2,$ 
because of the 
fact that creation-annihilation operators similarly to noncommutative $D_0$ brane field on YSTA are to 
be described by ${\cal D({\cal N})}\times {\cal D({\cal N})}$ matrices operating on ${\cal D({\cal N})}$
-dimensional Hilbert space I, according to our preceding postulate. This fact leads us to the important result that the Fock space or 
Hilbert space II constructed by a {\it finite} number of different creation-annihilation operators remains to be 
the ordinary Hilbert 
space with {\it countable} degrees of freedom. 

In contrast, it should be noted that 
in the ordinary local quantum field theory, there appear a countable, but {\it infinite} number of 
different creation-annihilation operators. They are introduced, for instance, by Fourier analysis 
applied to a local field \footnote{In this connection, 
it is interesting to note that in Ref. [7], we have derived the following $D_0$ brane field 
equation by applying the method of covariant Moyal star product to the $D_0$ brane field 
$D({\Sigma}_{aK})(= D(X_\mu, N))$ defined on YSTA, 
\begin{eqnarray}
[\ {\Sigma_{aK}}^2\ (\partial / {\partial \Sigma_{aL}})^2 - (\Sigma_{aK}  \partial/\partial {\Sigma_{aK}})^2 - 
(D-1) \Sigma_{aK}  \partial /{\partial \Sigma_{aK}}\ ]\ D (X_\mu, N)= 0,
\nonumber
\end{eqnarray}
with $K, L = b, \mu.$ It satisfies the translation invariance (see Eq. (3.1)) and tends to the massless 
Klein-Gordon equation in the 
quantum-mechanical limit, but is not fitted for a simple Fourier analysis, as was noticed there.}\ which is 
defined in the range $( R, - R)$ for each 
spatial coordinate and formally extended to an infinite spatial region by assuming an infinite 
repetition of the functional behavior in $(R, -R)$ or by taking the limit, $R \rightarrow \infty$.   
It is well-known that an {\it infinite} 
number of different creation-annihilation operators lead us to 
the {\it continuous} degrees of freedom of Fock space beyond the framework of the ordinary Hilbert space
which must be a fundamental origin of ultraviolet-divergence.

In addition, it should be noticed that the important concept of asymptotic field defined in the 
limit of time $t \rightarrow \pm \infty$ in the well-known Lehmann-Symanzik-Zimmermann's formalism 
of conventional local quantum field theory becomes inapplicable under the B-H-Z's proposition.
\vspace{1\baselineskip}
\section{\normalsize Concluding Remarks}

In the present paper, we have found that our postulate of a certain contraction of 
representation space of YSTA equipped with two scale parameters, $\lambda$ and $R$, in accordance with 
the B-H-Z's proposition leads us to the important result that the noncommutative field theory on 
YSTA applied to the actual physical world may be free from 
the ultraviolet divergence. It should be noticed here that, as 
seen from the argument given at the end in Sec. 4, this postulate becomes rather nominal 
and might be removable with fixed finite values of $\lambda$ and $R$ whenever our physical processes 
under consideration take place in accordance 
with the B-H-Z's proposition. In this case, however, one 
sees that the theoretical aspect of the finite spatial degrees of freedom of YSTA ceases to be explicit 
and instead its Lorentz covariance becomes explicit.

At this point, it is very interesting to ask how two scale parameters, $\lambda$ and $R$ in YSTA are to be 
fixed. 
According to the present view, it must be an experimental problem largely related to the applicability 
limit of the present quantum mechanics or local quantum field theory, because the canonical commutation 
relations in the ordinary quantum mechanics just appear as the quantum-mechanical limit of YSTA, that is,
the cooperative Inonu-Wigner's contraction limit, ${\cal N} (\equiv R /\lambda) \rightarrow \infty,$
 as was discussed in Sec. 3. It is doubtless that $\lambda$ might be 
identified with something like the Planck length $\l_P$ and $R$ with 
one of cosmological scales like $L$ in the B-H-Z' proposition. 

In this case, however, the 
relation of Yang's quantized space-time to the curved space-time or gravitation may be asked. 
As was discussed in Ref. [6], it is important to note again that YSTA which begins with the 
flat $(D+1,1)$-dimensional parameter space $q_M$'s subject to Eq. (2.2), might be regarded as a kind of flat local 
reference frame taken at any point in the $(D+1,1)$-dimensional curved space-time, on the 
analogy of the familiar local Lorentz frame in general theory of relativity, while the curved space-time 
must be self-consistent with the gravitation induced by the Matrix Model or $D_0$-brane field 
theory constructed on YSTA.

Several important subjects must be postponed to be discussed to the forthcoming paper, which should clarify 
the physical 
meaning of the reciprocity operator$^{[13]}$ $\hat{N}$ possibly underlying the relation between the present 
quantum mechanics and YSTA from the more profound level and study the Bekenstein-Hawking area-entropy 
relation in the light of our present scheme.

\end{document}